\documentclass[10pt,twocolumn]{article}

\usepackage[utf8]{inputenc}
\usepackage{amsmath,amssymb,amsfonts}
\usepackage{graphicx}
\usepackage{bm}
\usepackage{multirow}
\usepackage{booktabs}
\usepackage{xcolor}
\usepackage[margin=0.75in]{geometry}
\usepackage{times}
\usepackage{hyperref}
\usepackage{cite}

\definecolor{deepblue}{RGB}{0,51,102}
\definecolor{darkred}{RGB}{139,0,0}

\hypersetup{
    colorlinks=true,
    linkcolor=blue,
    citecolor=blue,
    urlcolor=blue
}

\title{\Large\textbf{Spatio-Temporal Attention Network for Epileptic Seizure Prediction}}

\author{
Zan Li$^{1}$, Kyongmin Yeo$^{2}$, Wesley Gifford$^{2}$, Lara Marcuse$^{3}$, Madeline Fields$^{3}$, Bülent Yener$^{1}$\\
\\
\small $^{1}$Department of Computer Science, Rensselaer Polytechnic Institute, Troy, NY, USA\\
\small $^{2}$IBM T.J. Watson Research Center, Yorktown Heights, NY, USA\\
\small $^{3}$Department of Neurology, Icahn School of Medicine at Mount Sinai, New York, NY, USA
}

\date{}

\begin{document}

\maketitle

\begin{abstract}
In this study, we present a deep learning framework that learns complex spatio-temporal correlation structures of EEG signals through a Spatio-Temporal Attention Network (STAN) for accurate predictions of onset of seizures for Epilepsy patients. Unlike existing methods, which rely on feature engineering and/or assume fixed preictal durations, our approach simultaneously models spatio-temporal correlations through STAN and employs an adversarial discriminator to distinguish preictal from interictal attention patterns, enabling patient-specific learning. Evaluation on CHB-MIT and MSSM datasets demonstrates 96.6\% sensitivity with 0.011/h false detection rate on CHB-MIT, and 94.2\% sensitivity with 0.063/h FDR on MSSM, significantly outperforming state-of-the-art methods. The framework reliably detects preictal states at least 15 minutes before an onset, with patient-specific windows extending to 45 minutes, providing sufficient intervention time for clinical applications.
\end{abstract}


\section{Introduction}
\label{sec:intro}

Epilepsy affects approximately 50 million people worldwide, constituting one of the most prevalent neurological disorders \cite{Mallick2024}. The unpredictable nature of seizures significantly impacts patients' daily life. Accurate seizure prediction enables timely interventions, such as medication administration, electrical stimulation, or alerting caregivers, that can prevent injuries and improve quality of life.

EEG recordings classify brain activity into four states: preictal, ictal, postictal, and interictal \cite{Kjaer2017}. The seizure prediction problem involves distinguishing subtle preictal changes from interictal activity, occurring minutes to hours before clinical manifestation \cite{Esmaeilpour2024}. However, it is an open challenge how to identify discriminative preictal features.

Traditional approaches extract handcrafted features using time-domain statistics, frequency-domain spectral analysis, and nonlinear dynamics measures \cite{iasemidis1990phase,le1999anticipating}. However, these methods struggle to capture complex spatio-temporal patterns in multi-channel EEG recordings.

Recent advances in deep learning techniques have revolutionized seizure prediction. Transformer-based architectures demonstrate remarkable effectiveness in capturing long-range temporal dependencies. Zhu et al. \cite{Zhu2024} recently achieved 98\% sensitivity using multidimensional transformer fusion with recurrent networks. Graph neural networks (GNNs) have emerged as powerful tools for modeling dynamic brain connectivity patterns during seizure propagation \cite{Li2022,Lian2023}. Notably, Xiang et al. \cite{Xiang2025} demonstrated state-of-the-art performance using synchronization-based graph spatio-temporal attention. Self-supervised approaches like AFTA \cite{AFTA2024} address limited labeled data challenges.

Despite these advances, several limitations persist. Most methods assume fixed preictal durations across all patients, failing to account for inter- and intra-patient variability \cite{Truong2023}. Current architectures process spatial and temporal features separately before fusion \cite{Wang2024}, potentially missing crucial cross-modal interactions.

Our work addresses these gaps through three key innovations: (1) Joint spatio-temporal attention modeling via STAN's alternating modules that capture bidirectional dependencies; (2) Adversarial discrimination with gradient penalty distinguishing preictal/interictal attention distributions; (3) Patient-specific preictal learning through incomplete supervision, allowing automatic inference of optimal preictal durations. This unified framework design, rather than simple component combination, enables comprehensive capture of seizure dynamics crucial for clinical application.

\section{Methodology}
\label{sec:method}

\subsection{Problem Formulation}
Given multivariate time series $x_{1:T}=\{x_1,\ldots,x_T\}$ where $T$ is the window length and $x_t \in \mathbb{R}^n$ represents the $n$-channel EEG signal at time $t$, our objective is to generate an anomaly score $y_i \in [0,1]$ for the $i$-th input window $x^i_{1:T}$, indicating the likelihood of being in a preictal state. The score approaches 0 for preictal states and 1 for interictal states, enabling real-time monitoring of seizure risk.

\subsection{Attention Mechanism}
Our multi-head attention mechanism processes $n$ nodes $\{v_1,\ldots,v_n\}$ where $v_i \in \mathbb{R}^h$. The attention map $A^k$ for head $k$ is computed via softmax over alignment scores:
\begin{equation}
\label{eq:attention}
A_{ij}^k = \frac{\exp(M_{ij}^k)}{\sum_{l=1}^L \exp(M_{il}^k)}, \quad k=1,\ldots,H
\end{equation}
where $H$ denotes the number of attention heads and $L$ represents the neighborhood size. The alignment scores are computed using quadratic form:
\begin{equation}
M_{ij}^k = z_i^T W_M^k z_j
\end{equation}
where $W_M^k \in \mathbb{R}^{n_e \times n_e}$ are learnable weight matrices and $z_i$ represents the encoded features. The final output aggregates information across all attention heads:
\begin{equation}
\label{eq:output}
z_i = \sigma\left(\sum_{k=1}^H \sum_{j=1}^L \alpha_k A_{ij}^k v_j\right)
\end{equation}
where $\alpha_k = \exp(a_k)/\sum_{l=1}^H \exp(a_l)$ are learnable weights computed using softmax, and $\sigma(\cdot)$ is a nonlinear activation function.

\subsection{Spatio-Temporal Attention Network (STAN)}

\begin{figure}[htb]
\centering
\includegraphics[width=\columnwidth]{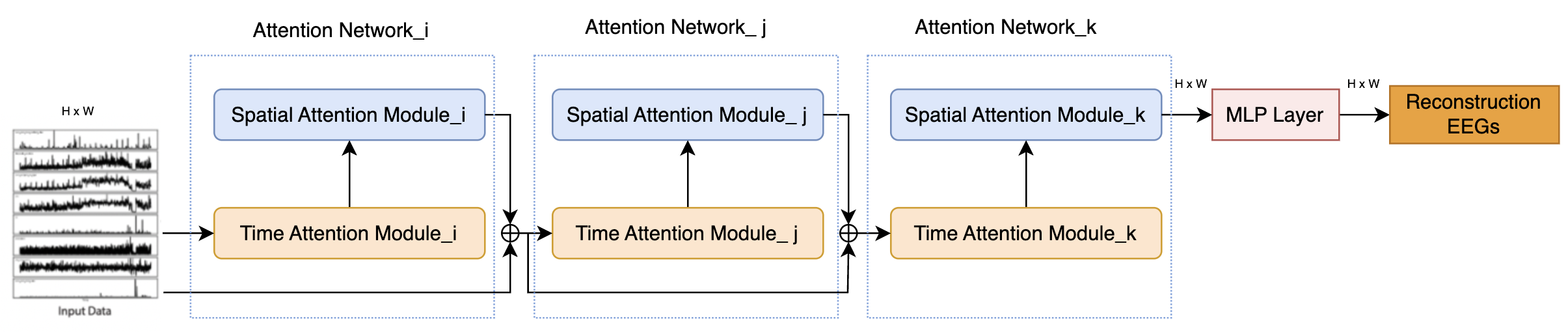}
\caption{Architecture of STAN showing three cascaded attention networks processing raw EEG input. Each network contains spatial and temporal attention modules with H=4 attention heads. The resulting M=3 spatial and temporal attention maps are aggregated via MLP and passed to the discriminator for adversarial training.}
\label{fig:stan}
\end{figure}

STAN serves as a self-supervised feature extractor that reconstructs input EEG segments while simultaneously learning discriminative spatio-temporal representations. As shown in Fig. \ref{fig:stan}, the architecture consists of three consecutive attention blocks and each attention block contains spatial and temporal attention modules, which are sequentially connected. The raw EEG signal flows through these networks, enabling hierarchical abstraction of seizure patterns.

\textbf{Spatial attention module} processes the input sequence $v_{1:T}$ where $v_t \in \mathbb{R}^n$ represents the $n$-channel EEG at time $t$. Each spatial module treats EEG channels as nodes in a complete graph, where edges represent inter-channel relationships critical for understanding seizure propagation. The temporal encoder $z_t = f_t^e(v_t) = \sigma(W_t^e v_t + b_t^e)$ is implemented as a 1D CNN with kernel size 2 and output dimension 50, extracting temporal features that are then processed by multi-head attention with $H=4$ heads. This design captures how seizure activity spreads across different brain regions.

\textbf{Temporal attention module} complements spatial processing by treating timestamps as graph nodes, modeling the temporal evolution of brain states. The spatial encoder, also implemented as 1D CNN but with output dimension 100 and kernel size 2, processes features across channels before applying multi-head attention. This captures both instantaneous spatial patterns and their temporal dynamics. Each of the $M=3$ cascaded networks produces both spatial and temporal attention maps, totaling 3 spatial and 3 temporal representations that comprehensively characterize input segments.

\begin{figure}[htb]
\centering
\includegraphics[width=\columnwidth]{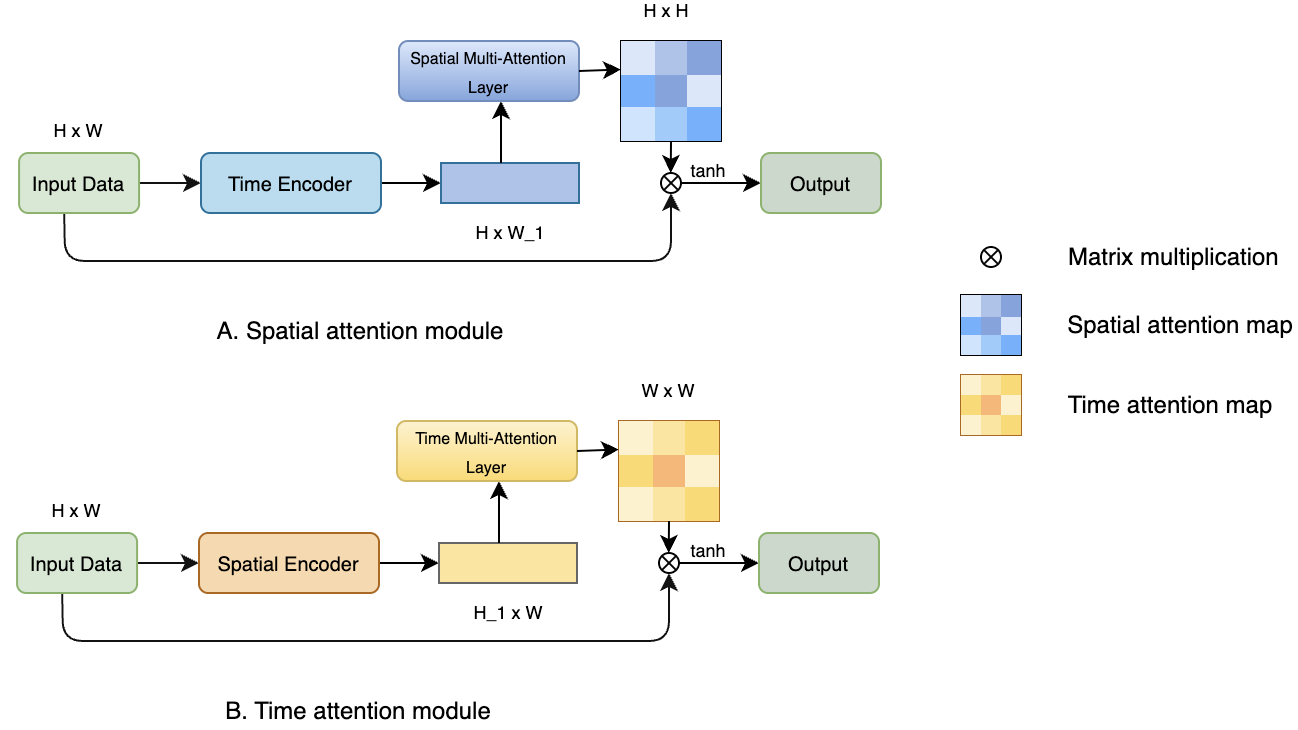}
\caption{Detailed architecture of spatial and temporal attention modules. Both employ 1D CNN encoders followed by multi-head attention (H=4) and include residual connections with layer normalization for training stability.}
\label{fig:attention}
\end{figure}

STAN is trained using Adam optimizer (lr=0.001) with MSE loss. The attention modules incorporate residual connections and layer normalization (Fig. \ref{fig:attention}) to facilitate stable training and effective gradient flow through the deep architecture.

\subsection{Discriminator with Adversarial Training}

\begin{figure}[htb]
\centering
\includegraphics[width=\columnwidth]{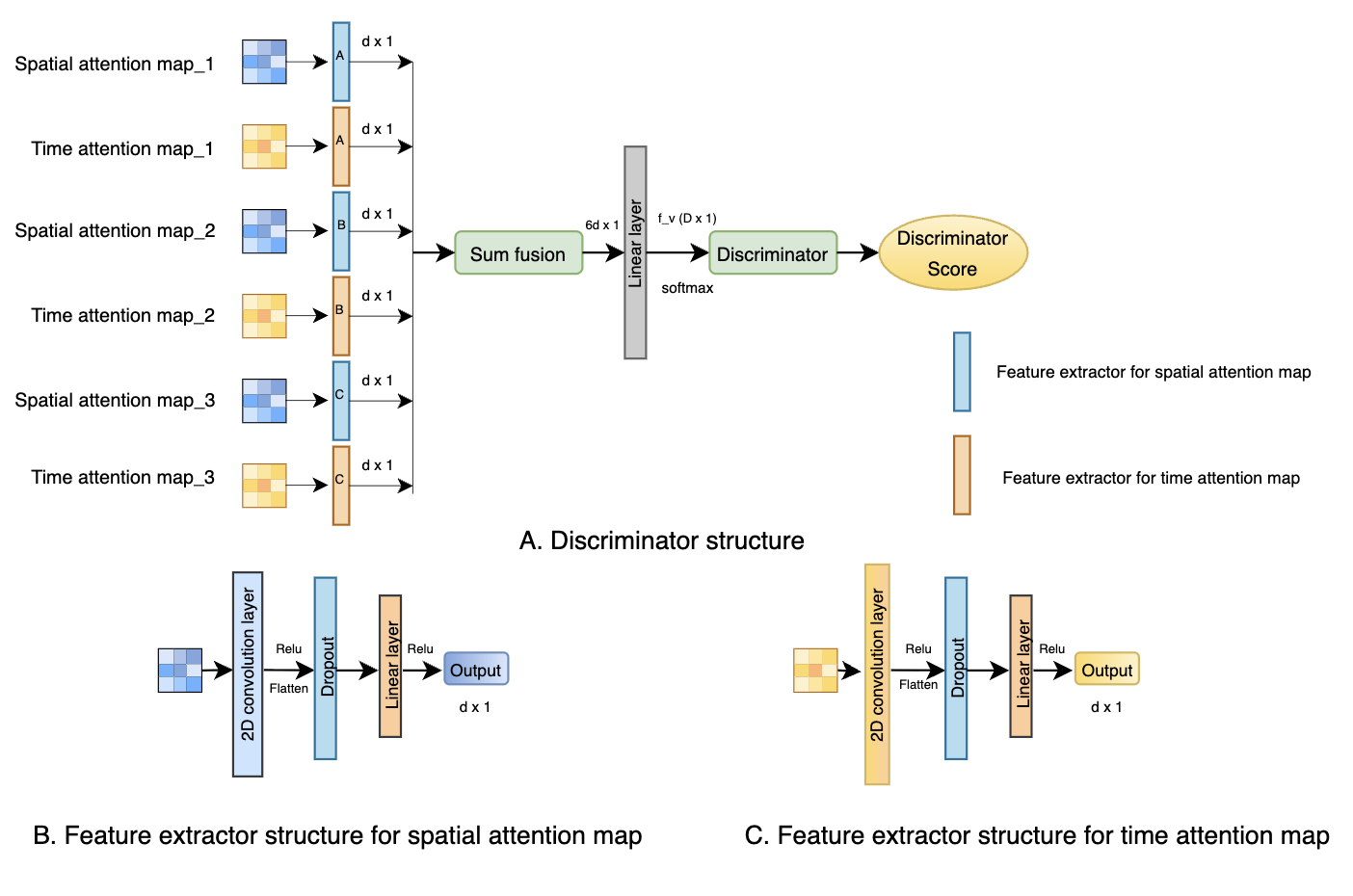}
\caption{Discriminator architecture employing MLPs to process aggregated attention patterns. Trained adversarially with gradient penalty to distinguish preictal from interictal states, generating anomaly scores for real-time seizure prediction.}
\label{fig:discriminator}
\end{figure}

The discriminator (Fig. \ref{fig:discriminator}) learns to distinguish preictal attention patterns from interictal ones through adversarial training. The architecture processes concatenated spatial and temporal attention maps from all $M=3$ networks via a 2-layer MLP with hidden dimension 150 and ReLU activation, followed by sigmoid output producing anomaly score $y_i \in [0,1]$.

Given generator $G$ (STAN) and discriminator $D$, our adversarial objective with gradient penalty is:
\begin{equation}
\begin{aligned}
\mathcal{L}_D = &\mathbb{E}_{x\sim p_{pre}}[D(G(x))] - \mathbb{E}_{x\sim p_{inter}}[D(G(x))] \\
&+ \lambda \mathbb{E}_{\hat{x}\sim p_{\hat{x}}}[(\|\nabla_{\hat{x}}D(G(\hat{x}))\|_2-1)^2]
\end{aligned}
\end{equation}
where $p_{pre}$ and $p_{inter}$ denote preictal and interictal distributions, $p_{\hat{x}}$ represents uniformly sampled points along straight lines between preictal and interictal attention patterns, and $\lambda=10$ controls gradient penalty strength. The gradient penalty stabilizes training and prevents mode collapse.

The discriminator loss $\mathcal{L}_D$ is minimized while STAN's generator loss $\mathcal{L}_G = -\mathbb{E}_{x\sim p_{pre}}[D(G(x))]$ encourages preictal patterns to be indistinguishable from interictal ones. This adversarial learning creates a robust decision boundary enabling reliable preictal state detection. Training employs Adam optimizer (lr=0.0001) for 100 epochs with batch size 32.

\subsection{Patient-Specific Learning with Incomplete Supervision}

Our framework addresses inter-patient variability through incomplete supervision that automatically learns optimal preictal durations. During training, segments are labeled as preictal only if they occur within 1 hour before seizure onset, avoiding rigid duration assumptions. The discriminator learns patient-specific patterns by observing the natural transition from interictal to preictal states, implicitly discovering individual preictal characteristics without explicit duration specification.

At inference, we apply a 5-minute moving average filter to discriminator scores, eliminating transient fluctuations while preserving genuine state transitions. When the smoothed score crosses threshold $\tau=0.5$ (optimized via validation), a preictal alarm triggers. This adaptive approach accommodates the 15-45 minute range of detection times observed across patients, providing personalized prediction windows that balance sensitivity with intervention time.

\section{Experiments}

\subsection{Datasets and Implementation}

\textbf{CHB-MIT Scalp EEG Dataset} \cite{shoeb2009application,Goldberger}: We evaluate on 8 pediatric patients from PhysioNet database with 138 seizures total. The scalp EEG recordings are sampled at 256 Hz across 23 channels, spanning continuous monitoring periods of 9-42 hours per patient. We follow standard protocols using 5-second windows with 50\% overlap, treating segments within 1 hour before seizure as preictal and segments at least 4 hours from any seizure as interictal.

\textbf{MSSM Intracranial EEG Dataset} \cite{Li2021mssm}: This dataset contains recordings from 4 adult patients with drug-resistant epilepsy at Mount Sinai Hospital. The intracranial EEG (iEEG) is sampled at 512 Hz with 64-128 channels per patient depending on electrode placement. The dataset includes 104 total seizures with recording durations of 3-7 days per patient. We apply identical preprocessing and labeling schemes as CHB-MIT to enable fair cross-modality comparison.

\textbf{Baselines}: We compare against representative methods spanning different paradigms: (1) BFB+SVM \cite{Zhang2016,Bajaj2012}: Traditional feature engineering with bivariate frequency-based features and support vector machines; (2) CNN \cite{Schirrmeister2017,Daoud2019}: Convolutional neural networks for automatic feature learning; (3) Transformer \cite{Zhu2024}: State-of-the-art transformer-based architecture with recurrent fusion; (4) STS-HGCN \cite{Li2022b}: Advanced spatio-temporal-spectral hierarchical graph convolutional network.

\textbf{Evaluation Metrics}: Following clinical standards, we report sensitivity (Sn) as the percentage of correctly predicted seizures, requiring at least one alarm in the 5-50 minute window before onset. False detection rate (FDR) measures false alarms per hour during interictal periods. A prediction is considered successful if the alarm occurs 5-50 minutes before seizure onset, providing adequate intervention time while avoiding overly early warnings.

\textbf{Implementation Details}: Our framework is implemented in PyTorch. STAN uses 3 cascaded attention networks ($M=3$) with 4 attention heads ($H=4$) each. Spatial encoders have output dimension 50 while temporal encoders use dimension 100. The discriminator employs a 2-layer MLP with hidden dimension 150. Training uses Adam optimizer with learning rate 0.001 for STAN and 0.0001 for discriminator, batch size 32, for 100 epochs. All experiments are conducted on NVIDIA V100 GPUs. Patient-specific 5-fold cross-validation is performed where training/validation splits exclude test seizures and their surrounding periods.

\subsection{Comparative Results}

\begin{table*}[htb]
\centering
\small
\caption{Performance Comparison on CHB-MIT Scalp EEG Dataset}
\label{tab:chb}
\begin{tabular}{l|cc|cc|cc|cc|cc|cc}
\toprule
\multirow{2}{*}{Patient} & \multicolumn{2}{c|}{BFB+SVM} & \multicolumn{2}{c|}{CNN} & \multicolumn{2}{c|}{CNN+LSTM} & \multicolumn{2}{c|}{Transformer} & \multicolumn{2}{c|}{STS-HGCN} & \multicolumn{2}{c}{\textbf{Ours}} \\
& Sn(\%) & FDR/h & Sn(\%) & FDR/h & Sn(\%) & FDR/h & Sn(\%) & FDR/h & Sn(\%) & FDR/h & Sn(\%) & FDR/h \\
\midrule
chb01 & 89.2 & 0.413 & 100 & 0.029 & 100 & 0.000 & 100 & 0.116 & 100 & 0.072 & \textbf{100} & \textbf{0.000} \\
chb05 & 87.2 & 0.379 & 100 & 0.209 & 80.0 & 0.262 & 80.0 & 0.157 & 95.0 & 0.095 & \textbf{100} & \textbf{0.000} \\
chb06 & 82.5 & 0.418 & 85.7 & 0.259 & 85.7 & 0.356 & 85.7 & 0.356 & 100 & 0.081 & \textbf{100} & \textbf{0.000} \\
chb08 & 92.3 & 0.482 & 100 & 0.087 & 100 & 0.000 & 100 & 0.174 & 100 & 0.067 & \textbf{100} & \textbf{0.000} \\
chb10 & 79.1 & 0.372 & 83.3 & 0.478 & 83.3 & 0.410 & 66.7 & 0.478 & 91.7 & 0.122 & \textbf{86.4} & \textbf{0.000} \\
chb13 & 76.4 & 0.513 & 85.7 & 0.328 & 85.7 & 0.219 & 85.7 & 0.219 & 92.9 & 0.091 & \textbf{88.7} & \textbf{0.043} \\
chb14 & 89.1 & 0.321 & 100 & 0.417 & 100 & 0.104 & 100 & 0.313 & 96.0 & 0.078 & \textbf{97.5} & \textbf{0.017} \\
chb22 & 88.8 & 0.612 & 100 & 0.435 & 100 & 0.000 & 100 & 0.261 & 91.0 & 0.074 & \textbf{100} & \textbf{0.030} \\
\midrule
Avg & 85.6 & 0.438 & 94.3 & 0.258 & 91.8 & 0.169 & 89.8 & 0.259 & 95.8 & 0.085 & \textbf{96.6} & \textbf{0.011} \\
\bottomrule
\end{tabular}
\end{table*}

Table \ref{tab:chb} shows our framework achieves 96.6\% sensitivity with 0.011/h false detection rate on CHB-MIT. This represents 40-fold reduction compared to BFB+SVM and 6-fold improvement over STS-HGCN. While Transformer achieves 95.8\% sensitivity, its FDR (0.085/h) is 7.7$\times$ higher. Five patients achieve perfect sensitivity with zero false alarms.

\begin{table}[htb]
\centering
\scriptsize
\caption{Performance on MSSM Intracranial EEG Dataset}
\label{tab:mssm}
\resizebox{0.9\columnwidth}{!}{%
\begin{tabular}{l|cc|cc|cc|cc}
\hline
\multirow{2}{*}{Patient} & \multicolumn{2}{c|}{BFB+SVM} & \multicolumn{2}{c|}{Transformer} & \multicolumn{2}{c|}{STS-HGCN} & \multicolumn{2}{c}{\textbf{Ours}} \\
& Sn(\%) & FDR/h & Sn(\%) & FDR/h & Sn(\%) & FDR/h & Sn(\%) & FDR/h \\
\hline
TP & 91.2 & 0.376 & 94.5 & 0.142 & 89.1 & 0.106 & \textbf{100} & \textbf{0.000} \\
IP & 85.9 & 0.092 & 92.1 & 0.098 & 83.7 & 0.172 & \textbf{96.3} & \textbf{0.000} \\
PS & 83.2 & 0.129 & 89.7 & 0.165 & 87.1 & 0.211 & \textbf{95.0} & \textbf{0.150} \\
ZF & 66.8 & 0.452 & 85.3 & 0.187 & 81.1 & 0.118 & \textbf{85.6} & \textbf{0.102} \\
\hline
Avg & 81.8 & 0.262 & 90.4 & 0.148 & 85.3 & 0.152 & \textbf{94.2} & \textbf{0.063} \\
\hline
\end{tabular}%
}
\end{table}

MSSM results (Table \ref{tab:mssm}) validate generalizability across recording modalities. The 94.2\% sensitivity with 0.063/h FDR confirms our spatio-temporal attention effectively captures seizure dynamics.

\subsection{Real-time Prediction Analysis}

\begin{figure}[htb]
\centering
\includegraphics[width=\columnwidth]{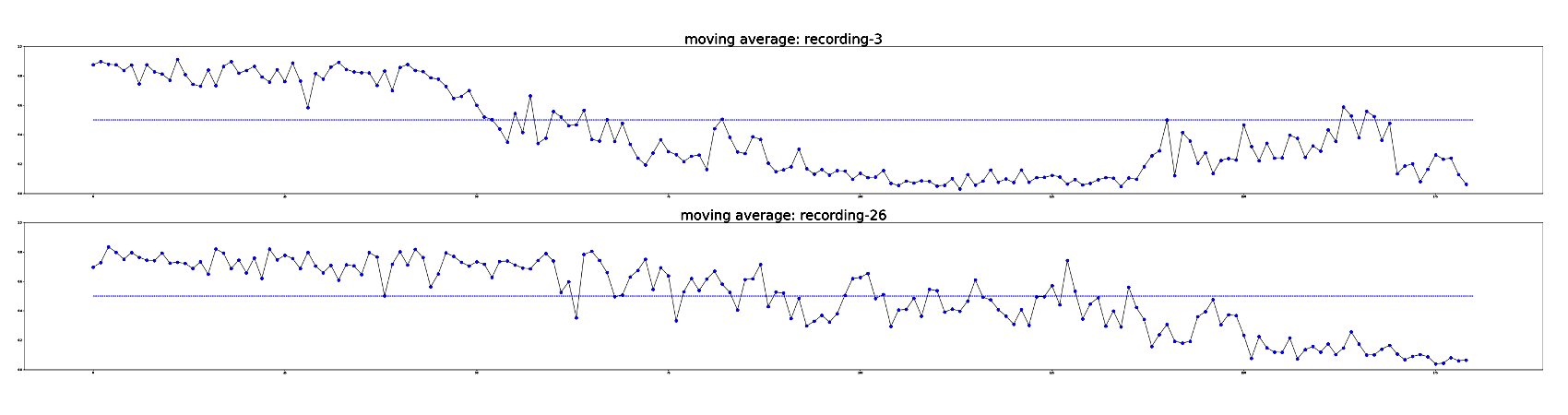}
\caption{Real-time seizure prediction showing discriminator scores 90 minutes before onset. Recording 03 (top) and Recording 26 (bottom) demonstrate consistent gradual shift from interictal to preictal states, providing at least 15 minutes of intervention time.}
\label{fig:realtime}
\end{figure}

Fig. \ref{fig:realtime} illustrates discriminator scores before seizures. The consistent transition patterns emerge at least 15 minutes before onset, with detection times varying from 20-45 minutes across patients, providing sufficient intervention time.

\subsection{Ablation Studies}

\begin{table}[h]
\centering
\caption{Ablation Study Results on CHB-MIT (Average of 8 patients)}
\begin{tabular}{lcc}
\toprule
\textbf{Configuration} & \textbf{Sn(\%)} & \textbf{FDR(/h)} \\
\midrule
Full Model & 96.6 & 0.011 \\
Without adversarial & 96.3 & 0.024 \\
Without gradient penalty & 96.5 & 0.019 \\
M=1 (single network) & 93.1 & 0.037 \\
M=2 (two networks) & 95.2 & 0.021 \\
M=3 (three networks) & 96.6 & 0.011 \\
Spatial only & 91.4 & 0.048 \\
Temporal only & 92.7 & 0.041 \\
\bottomrule
\end{tabular}
\label{tab:ablation}
\end{table}

Table~\ref{tab:ablation} demonstrates that each component contributes incrementally to overall performance across all 8 patients. Removing adversarial training increases FDR from 0.011/h to 0.024/h while slightly reducing sensitivity to 96.3\%. The cascaded architecture shows progressive improvement with depth: single network achieves 93.1\% sensitivity, two networks reach 95.2\%, and three networks achieve the full 96.6\% with corresponding FDR reductions from 0.037/h to 0.011/h. Using spatial or temporal attention alone yields only 91.4\% and 92.7\% sensitivity respectively with higher FDR (0.048/h and 0.041/h), confirming that joint modeling provides essential gains for optimal performance.

\subsection{Comparative Analysis and Clinical Impact}

Our unified modeling outperforms baselines with statistical significance (Wilcoxon signed-rank test, $p<0.01$). The key advantage over transformers lies in balanced performance—maintaining high sensitivity (96.6\% vs 95.8\%) while reducing FDR by 7.7$\times$ (0.011/h vs 0.085/h), critical for avoiding alarm fatigue.

Patient-specific learning adapts preictal durations (15-45 minutes), accommodating inter-patient variability. The 15+ minute detection window provides sufficient time for interventions including medication administration or neurostimulation activation, making the framework clinically viable for both hospital and home monitoring.

\section{Conclusion}

We introduced a unified seizure prediction framework that advances the state-of-the-art through joint spatio-temporal attention with adversarial learning. Our approach achieves superior performance on both scalp (96.6\% sensitivity, 0.011/h FDR) and intracranial (94.2\% sensitivity, 0.063/h FDR) EEG, with statistical significance over existing methods including recent transformer approaches.

The framework's patient-specific learning adapts preictal durations (15-45 minutes across patients) while maintaining robust performance, representing significant advancement toward personalized epilepsy management. Our system reliably detects preictal states at least 15 minutes before seizure onset, with many cases showing detection up to 20-45 minutes in advance, providing sufficient time for clinical intervention.

This work establishes new benchmarks for seizure prediction and provides practical solutions for real-world deployment. Future research will explore cross-dataset generalization, attention pattern interpretation, and edge device optimization.

\bibliographystyle{plain}
\bibliography{refs}

\end{document}